\documentclass{camera}
\usepackage{epsfig}  
\usepackage{physics}

\def\SLC{{\sc SLC}}
\def\SLD{{\sc SLD}}
\def\LEP{{\sc LEP}}
\def\Af{\ensuremath{A_{\mathrm{f}}}}
\def\Ae{\ensuremath{A_{e}}}
\def\Al{\ensuremath{A_{\ell}}}
\def\ptau{\ensuremath{\cal P_{\tau}}}
\def\Ab{\ensuremath{A_{\mathrm{b}}}}
\def\Ac{\ensuremath{A_{\mathrm{c}}}}

\begin{document}

%
\title{LEPTON AND QUARK ASYMMETRIES}

%
\author{V. Ciulli}

%
\organization{Istituto Nazionale di Fisica Nucleare, Sezione di Firenze}

%

\maketitle

\begin{abstract}
Lepton and quark asymmetries at Z boson peak are a powerful 
means to test the couplings of the Z boson to fermions. The measurements 
performed at \LEP\ and \SLC\ colliders are
reviewed, and their impact on the determination of the electroweak
mixing angle is presented. 
\end{abstract}

\section{Asymmetry parameters}

The asymmetries of fermions in $\ee$ collisions at the Z
peak are related to the vector and axial-vector couplings of the Z
boson to fermions, ${\gv}_\mathrm{f}$ and ${\ga}_\mathrm{f}$, or, more
precisely, to the asymmetry paramater 
$$\Af=\frac{2{\gv}_\mathrm{f}{\ga}_\mathrm{f}}{{\gv}_\mathrm{f}^2+{\ga}_\mathrm{f}^2}\,
  .$$ In particular, the left-right asymmetry, $\ALR$, i.e.   
the difference between the cross-sections for 
left/right polarised electrons divided by the total cross-section, 
is equal to $\Ae$. For unpolarised beams instead, the polarisation of
the outgoing fermions f, ${\cal P}_\mathrm{f}$, is equal to $-\Af$.  
Finally the forward-backward asymmetry, $\AFB^\mathrm{f} =
  \frac{3}{4}\Ae\Af$, can be measured from the distribution of the angle between the direction
  of the outgoing fermion f and the incoming electron, and it
  is the only asymmetry which does not require a polarisation
  measurement. From each of these measurements the parameter $\swel$,
  related to the weak mixing angle
of the Standard Model, can be extracted.

\section{Measurements at \LEP\ and \SLC.}
A total of 17 millions Z decays into hadrons and leptons have been
collected altogether by the four experiments at \LEP\ collider between 1990
and 1995. This huge statistics allowed to
measure forward-backward asymmetries of all leptons and of heavy
quarks, and the polarisation of the $\tau$ lepton~\cite{lepew}. 
Much less events, about 500000 Z bosons, were collected in 1993-1998 by the {\sc SLD}
experiment at the \SLC\ collider, but by means of the 80\%
longitudinally polarised electron beam, they allowed a very precise measurement
of $\ALR$ and of $\Ab$~\cite{lepew}.   

If lepton universality is assumed, 
the parameter \Al\ can be measured from \ALR, the
forward-backward lepton asymmetries,
$\AFB^\ell$, and the $\tau$ polarisation, $\ptau$.
Results from these measurements are shown in Table~\ref{al} and they
are all in agreement.

\begin{table}[b]
\caption{Results for \Al\ from \LEP\ (top) and \SLD\ (bottom).}
$$
\begin{array}{|c|c|c|c|}\hline
 & \Al & \mathrm{Cumulative\, average} & \chi^2/\mathrm{d.o.f.} \\ \hline
\AFB^\ell & 0.1512\pm0.0042& & \\
\ptau & 0.1465\pm0.0033 &0.1482\pm0.0026 & 0.8/1\\ \hline
\ALR & 0.1513\pm0.0021& 0.1501\pm0.0016& 1.6/2\\
\hline
\end{array}
$$
\label{al}
\end{table}


The parameters \Ab\ and \Ac\ are measured directly by \SLD\ from the
corresponding $\AFB$'s with left and right polarised electron beam. It
is, however, possible to extract 
them from $\AFB^\mathrm{b}$ and $\AFB^\mathrm{c}$  measured at \LEP,
and \Al. 
\begin{table}
\caption{Results for \Ab\ and \Ac.}
$$
\begin{array}{|c|c|c|c|c|}\hline
& \mathrm{\LEP} & \mathrm{\SLD} & \mathrm{\LEP} + \mathrm{\SLD} &
  \mathrm{SM} \\  
& \Al=0.1482\pm0.0026 &  & \Al=0.1501\pm0.0016 & \mathrm{fit} \\ \hline
\Ab & 0.894\pm0.022& 0.922\pm0.020 & 0.901\pm0.013 & 0.935 \\
\Ac & 0.639\pm0.034& 0.670\pm0.026 & 0.656\pm0.021 & 0.668 \\
\hline
\end{array}
$$
\label{ab}
\end{table}
Results for \Ab\ and \Ac\ are shown in Table~\ref{ab}, and compared to 
the predictions from the Standard Model fit.  
A discrepancy by 2.6 standard deviations is observed between the measured 
\Ab\ and the Standard Model prediction, when combining \LEP\ and
\SLD\ results. No discrepancies are observed instead for \Ac. 
\begin{figure}
\center{\epsfig{figure=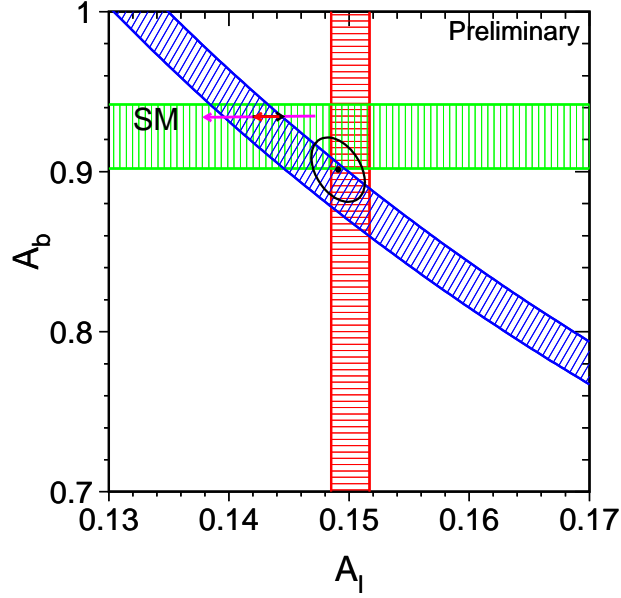,height=7.9cm}}
\caption{Comparison of the measurements of \Ab, $\AFB^\mathrm{b}$ and
  \Al. Bands of $\pm1$ standard deviation are shown, together with
  the 68\% confidence level contour for the joint analysis. The 
  arrows pointing to the right and to the left show the variation in
  the Standard Model prediction for varying $m_\mathrm{t}$ in the range
  $174.3\pm5.1$ \GeVcc\ and $m_\mathrm{H}$ in the range
  $300^{+700}_{-186}$ \GeVcc, respectively.}
\label{alab} 
\end{figure}
In Fig.~\ref{alab} the results are shown in the (\Al,\Ab) plane. 
The measured $\AFB^\mathrm{b}$ is compatible
with \Ab\ measured by \SLD\ and the Standard Model, but 
the resulting prediction for \Al\ is significantly lower than the measured value.


\begin{figure}[h]
\center{\epsfig{figure=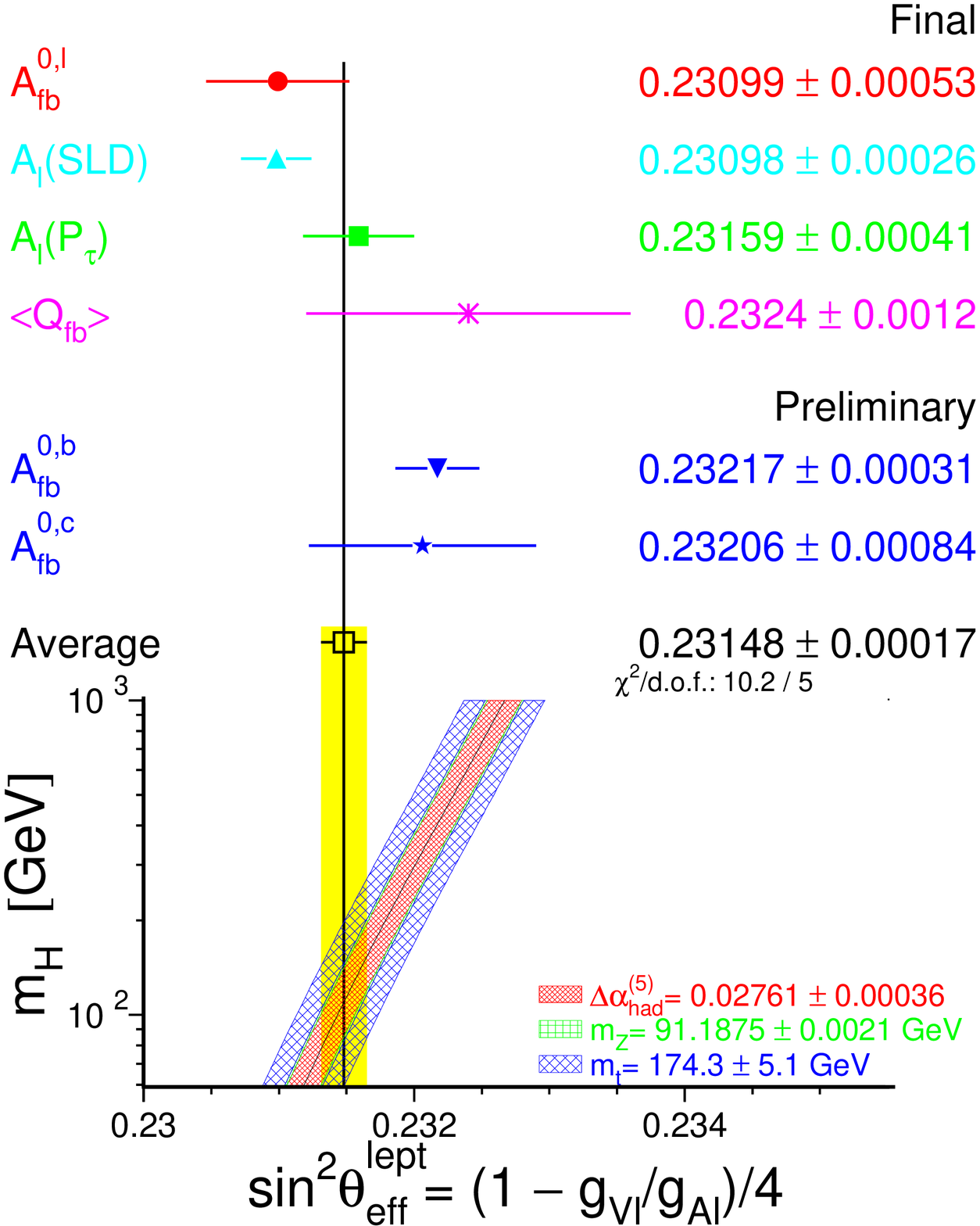,height=7.9cm}}
\caption{Results for \swel. The Standard Model prediction 
is also shown.}
\label{sef} 
\end{figure}

The combination of all the  
asymmetry measurements yields $\swel=
0.23148\pm0.00017$. But, as shown in Fig.~\ref{sef}, the
most precise results, those from \ALR\ and $\AFB^\mathrm{b}$, differ
by about 2.9 standard deviations. Altogether, the $\chi^2/$d.o.f. of
the fit is 10.2/5, which has 7\% probability to occur.

\section{Conclusions}

An impressive precision has been reached on $\swel$ from the
asymmetry measurements at the Z peak. A three sigma discrepancy,
however, is still present between $\ALR$ and $\AFB^\mathrm{b}$. 
In general lepton measurements are in agreement with each other
but not with quark measurements, even though the discrepancy would be below 
two sigmas if \LEP\ measurements only were considered. 
A deviation of the Z couplings to the quarks 
which shows up only in these measurements is
unlikely. Only a large systematic effect, either common to all \LEP\
experiments or much larger than the estimated systematic uncertainty
on \ALR, could explain such a difference, if it is not a statistical 
fluctuation.  Finally, it is worth noticing that only few 
measurements are still preliminary, therefore it is likely that this
discrepancy will not be solved before the next generation of
experiments at future colliders.   


%
\end{document}